\documentclass[useAMS,usenatbib]{mn2e}
\usepackage{psfig}

\title[Linking quasars to dark matter halos]
{A simple model to link the properties of quasars to the properties of dark matter halos out to high redshift}

\author[D.~J.~Croton]{
\parbox[t]{\textwidth}{
Darren J. Croton$^{1,2}$
}
\vspace*{6pt} \\ 
$^1$Department of Astronomy, University of California, Berkeley, CA, 94720, USA\\
$^2$Centre for Astrophysics \& Supercomputing, Swinburne University of Technology, P.O. Box 218, Hawthorn, VIC 3122, Australia
\vspace{-0.5cm} 
}

\date{Accepted ---. Received ---;in original form ---}

\pubyear{2005}

\newcommand{\plotone}[1]
           {\centering \leavevmode \psfig{file=#1,width=\columnwidth,clip=}}

\def\simlt{\lower.5ex\hbox{$\; \buildrel < \over \sim \;$}}
\def\simgt{\lower.5ex\hbox{$\; \buildrel > \over \sim \;$}}

\begin{document}

\maketitle


\begin{abstract}
We present a simple model of how quasars occupy dark matter halos from $z=0$ to $z=5$ using the observed $m_{\rm BH}-\sigma$ relation and quasar luminosity functions.  This provides a way for observers to statistically infer host halo masses for quasar observations using luminosity and redshift alone.  Our model is deliberately simple and sidesteps any need to explicitly describe the physics.  In spite of its simplicity, the model reproduces many key observations and has predictive power: 1) model quasars have the correct luminosity function (by construction) and spatial clustering (by consequence); 2) we predict high redshift quasars of a given luminosity live in less massive dark matter halos than the same luminosity quasars at low redshifts; 3) we predict a factor of $\sim 5$ more $10^{8.5}M_\odot$ black holes at $z\sim2$ than is currently observed; 4) we predict a factor of $\sim 20$ evolution in the amplitude of the $m_{\rm BH}-M_{\rm halo}$ relation between $z=5$ and the present day; 5) we expect luminosity dependent quasar lifetimes of between $t_Q\sim 10^{7-8}\,{\rm yr}$, but which may become as short as $10^{5-6}\,{\rm yr}$ for quasars brighter than $L^*$; 6) while little luminosity dependent clustering evolution is expected at $z \simlt 1$, increasingly strong evolution is predicted for $L>L^*$ quasars at higher redshifts.  These last two results arise from the narrowing distribution of halo masses that quasars occupy as the Universe ages. We also deconstruct both ``downsizing'' and ``upsizing'' trends predicted by the model at different redshifts and space densities. Importantly, this work illustrates how current observations cannot distinguish between more complicated physically motivated quasar models and our simple phenomenological approach.  It highlights the opportunities such methodologies provide.
\end{abstract}

\begin{keywords}
	quasars: general, galaxies: active, cosmology: dark matter, methods: statistical
\end{keywords}

\section{Introduction}
\label{sec:intro}

Quasars represent a unique population of objects in the Universe that encapsulate many otherwise diverse areas of physics.  These include extreme environments of gravity (black holes), sub-to-kiloparsec-scale dynamics (black hole two and three body interactions and host galaxy mergers or secular triggers of quasar activity), sub-to-kiloparsec-scale hydrodynamics (gas infall, accretion disks and quasar winds), and quasars as cosmological probes of the evolving large-scale cosmic web.  They are among the most luminous objects in the Universe.  The energy liberated during a single quasar event can outshine the entire stellar light of the host galaxy.  After fading, their presence can still be measured through the local quiescent black hole population.

Although much work has been done to describe the physics of black holes and their evolution, the majority of what we know remains primarily phenomenological. Two fundamental correlations are observed: the $m_{\rm BH}-\sigma$ relation and the $m_{\rm BH}-m_{\rm bulge}$ relation \citep{Ferrarese2000, Tremaine2002, Marconi2003, Haring2004}.  Although initially surprising (why should the sub-parsec physics of black hole growth correlate with the kiloparsec properties of the galactic bulge?), it is now believed that these relationships simply reflect the physics of a common formation mechanism \citep{Silk1998, Hopkins2006a}.  For example, galaxy major mergers may simultaneously drive growth in the bulge and force gas into the central regions of the galaxy to fuel the black hole (and hence a quasar).  Secular processes may be operating to similar effect, such as bar instabilities \citep{Sellwood1999}.

There is a need to understand the phenomenology of black holes and galaxies in greater detail.  Active black holes are thought to have great impact on the evolution of their hosts.  During a quasar event, rapid hole growth occurs and outflows drive winds that may liberate gas from the galaxy \citep{DiMatteo2005, Hopkins2006a, Thacker2006}.  Such gas is considered fuel for future star formation, and without it new stars no longer form and the host galaxy subsequently reddens and fades, a kind of galactic extinction.  However, supermassive black holes do not always die with their galaxy, but are often later found in a low luminosity state.  Heating from low luminosity active galactic nuclei (AGN) provide a long term energy source that suspends the cooling of gas from the surrounding hot halo \citep[the so called ``radio mode'' solution to the ``cooling flow'' problem:][]{Croton2006, Bower2006}.  This heating/cooling balance is known to maintain the aged appearance of many massive local galaxies.  Understanding AGN and black holes has now become essential to understanding galaxies, and hence modelling the co-evolution of both has become a subject of great interest.

In this paper we build a general model of how black holes and quasars occupy dark matter halos and how this occupation evolves with time.  Our method is similar in spirit to that by \cite{Marinoni2002} and \cite{Vale2004} but for quasars rather than galaxies.  We work from a minimal set of assumptions and use two key observational constraints: the $m_{\rm BH}-\sigma$ relation and the quasar luminosity function.  Under such constraints the model naturally reproduces many quasar and black hole observations out to redshifts as distant as $z\sim5$.  A number of predictions are given.  This model is deliberately simple; it sidesteps any attempt to explicitly describe the quasar triggering mechanism, the details of black hole accretion, or the hydrodynamics of subsequent quasar winds and outflows.  Although understanding such detail is certainly desirable, we show that the current observations do not discriminate between the more complicated physically motivated modelling of quasars and our simple phenomenological model.

The outline of this paper is as follows.  In Section~\ref{sec:method} we describe the construction of the model and the observational data used to constrain it.  In Section~\ref{sec:results} we explore the various consequences of this model, comparing to observations where available and making predictions where not.  Section~\ref{sec:discussion} provides some discussion, placing the model into a broader context of black hole and galaxy co-evolution.  Finally, Section~\ref{sec:summary} summarises our main results.  Unless otherwise stated, we assume a standard WMAP first year $\Lambda$CDM cosmology with $\Omega_\Lambda=0.75$, $\Omega_{\rm m}=0.25$, $\sigma_8=0.9$ and Hubble constant $H_0=100\,h\,{\rm km/s}\,{\rm Mpc}^{-1}$ \citep{Spergel2003, Seljak2005}.  We choose the value of the Hubble parameter, $h$, to be either $h=0.7$ or $h=1.0$ depending on the context; this will be clearly marked.

\section{Accurately populating dark matter halos with quasars}
\label{sec:method}

We build our quasar model in two parts.  First, we map quasar luminosity onto dark matter halos using the halo virial properties.  Second, we determine which of these halos actually host a quasar as a function of luminosity at any given redshift.  Both parts are undertaken using observational constraints only, notably the $m_{\rm BH}-\sigma$ relation and quasar luminosity function.

To begin we require knowledge of the dark matter halo population and its evolution.  There are several ways this can be achieved at a given redshift for a given cosmology.  Analytic methods, such as those recently described in \cite{Neistein2008} and \cite{Zhang2008}, are not as useful here as they produce halo merger trees lacking spatial and velocity information.  We will later need these properties in our analysis.

Instead, we turn to a numerical N-body simulation of dark matter evolution, the Millennium Simulation \citep{Springel2005}.  This simulation, run using a WMAP1+2DFGRS cosmology, follows the evolution of 10 billion dark matter particles in a box of side-length $500\,h^{-1}\,{\rm Mpc}$ from $z=127$ to $z=0$.  Within the simulation, both halos and subhalos (i.e. the bound sub-structure within a given halo) are accurately resolved down to virial masses of less than $10^{11}\,M_\odot$, more than sufficient for our purposes here.  Note that in what follows we do not discriminate between halos and subhalos when populating the simulation with quasars\footnote{Throughout this paper our use of the term ``halo'' includes both halos (i.e. quasars within central galaxies) and subhalos (i.e. quasars within satellite galaxies).}.  Thus, it is possible in our model for given halo to host more than one quasar at any given time.  Small-scale clustering measures of quasars indicate that this may indeed be the case \citep[e.g.][]{Hennawi2006, Myers2008, Padmanabhan2008}, although it is not critical for our current analysis.

\subsection{Linking quasar luminosity with halo mass}
\label{sec:mapping}

We start with the dark matter halo virial mass, $M_{\rm vir}$, of each Millennium Simulation halo.  Our goal is to first relate its properties to the expected velocity dispersion of an occupying galaxy, $\sigma$, and then to the quasar luminosity through the $m_{\rm BH}-\sigma$ correlation.  Readers who are only interested in the final relations should skip ahead to the equation summary in Section~\ref{sec:eqns} (see also equation~\ref{eqn:mag2mass} in Section~\ref{sec:application}).

Dark matter halos are identified at each redshift as regions of the simulation whose mean density inside a spherical aperture exceeds 200 times the critical density of the Universe.  The virial mass of a halo and its virial velocity, $V_{\rm vir}$, are then related by
\begin{equation}
	V_{\rm vir} = [ 10\,G\,H(z)\,M_{\rm vir} ]^{1/3}~,
	\label{eqn:vvir}
\end{equation}
where $M_{\rm vir}$ has units of $h^{-1} M_\odot$, $G=4.3\times 10^{-9}\,({\rm km/s})^2\,{\rm Mpc}\,M_\odot^{-1}$ is Newtons gravitational constant, and
\begin{equation}
	H(z) \equiv H_0\, E(z) = H_0\, [\Omega_{\rm m} (1+z)^3 + \Omega_\Lambda]^{1/2}
	\label{eqn:Hz}
\end{equation}
is the value of the Hubble constant at redshift $z$  \citep{Hogg1999}.  Here, $H_0=100\,h\,{\rm km/s}\,{\rm Mpc}^{-1}$ is the local value of the Hubble constant (with $h$ the dimensionless Hubble value), while $\Omega_{\rm m}$ and $\Omega_\Lambda$ are the universal mass and dark energy densities. Note, that both $M_{\rm vir}$ and $H(z)$ above must share the same $h$ value for dimensional consistency.

Halo virial velocity is typically related to the observed galaxy circular velocity, $v_c$, by
\begin{equation}
	v_c = \gamma\, V_{\rm vir}~,
	\label{eqn:vc}
\end{equation}
with $\gamma$ a parameter of order unity (see Section~\ref{sec:assumtions}).  This in turn can be related to the velocity dispersion of the galaxy, $\sigma$, through the observed correlation \citep{Baes2003}
\begin{equation}
	\log \big(\frac{\sigma}{u_0}\big) = (-0.22 \pm 0.03) + (1.04 \pm 0.12) \log \big(\frac{v_c}{u_0}\big)~,
	\label{eqn:sigma}
\end{equation}
where $u_0=200\,{\rm km/s}$.  Equations~\ref{eqn:vvir}-\ref{eqn:sigma} connect the virial mass of a dark matter halo at redshift $z$ to the velocity dispersion of the occupying galaxy.

A well defined correlation between stellar velocity dispersion and black hole mass is observed in the local Universe, the $m_{\rm BH}-\sigma$ relation \citep{Tremaine2002}:
\begin{equation}
	\log \big(\frac{m_{\rm BH}}{h_{70}^{-1}\, M_\odot}\big) = (8.13 \pm 0.09) + (4.02 \pm 0.44) \log \big(\frac{\sigma}{\sigma_0}\big),
	\label{eqn:msigma}
\end{equation}
where $h_{70}$ is the Hubble parameter with $H_0=70\,{\rm km/s}\,{\rm Mpc}^{-1}$, and $\sigma_0=200 {\rm km/s}$.  For practical purposes, when using equation~\ref{eqn:msigma} in our model we include the observed $0.3\,{\rm dex}$ dispersion in $m_{\rm BH}$ (for simplicity we apply the same dispersion at all values of $\sigma$).  We assume no evolution with redshift in either amplitude or slope of the relation.  We discuss this further in Section~\ref{sec:msigma}.

With a value of $m_{\rm BH}$ for each dark matter halo we can now determine the luminosity of a quasar that may occupy the halo.  We take this (bolometric) luminosity as some fraction, $\eta$, of the Eddington luminosity:
\begin{eqnarray}
	L_Q & = & \eta\, L_{\rm edd} \nonumber \\
	& = & \eta\ 3.3\times 10^4\, m_{\rm BH}\, (h_{70}^{-1}\, L_\odot) \nonumber \\
	& = & \eta\ 1.3\times 10^{38}\, m_{\rm BH}\, (h_{70}^{-1}\, {\rm erg/s})~.
	\label{eqn:LQ}
\end{eqnarray}
Equation~\ref{eqn:LQ} completes our sought after connection between virial mass and quasar luminosity, which we will now summarise.  

\subsubsection{A summary of the important equations and relationships}
\label{sec:eqns}

Equations~\ref{eqn:vvir}-\ref{eqn:LQ} provide a mapping between halo virial mass and quasar bolometric luminosity.  These equations can be reduced to the following relation
\begin{eqnarray}
	\lefteqn{\log\Big( \frac{L_Q/\eta}{10^{12}\,h_{70}^{-1}\,L_\odot} \Big)\ = \ (-1.99\pm0.33) } \nonumber \\
	& & + \ (1.39\pm0.22)\,\log \Big [ \gamma^3\, H(z) \Big( \frac{M_{\rm vir}}{10^{13}\,h^{-1}\,M_\odot} \Big) \Big]~,
	\label{eqn:mass2mag}
\end{eqnarray}
where $H(z)$ is defined by equation~\ref{eqn:Hz}.  Note that, for internal consistency, $H(z)$ must be calculated using the same value of $h$ as $M_{\rm vir}$, but for $L_Q$ we have assumed the standard observer value of $h=0.7$. Quasar luminosity here is dependent on both parameters $\gamma$, describing the relationship between virial and circular velocities (equation~\ref{eqn:vc}), and $\eta$, the Eddington luminosity fraction that quasars are assumed to shine at (equation~\ref{eqn:LQ}).  Observational errors from equations~\ref{eqn:sigma} and \ref{eqn:msigma} have been propagated throughout; their impact on the results are discussed in Section~\ref{sec:errors}.

Equations~\ref{eqn:vvir}-\ref{eqn:msigma} can also be reduced to map directly between halo mass and black hole mass
\begin{eqnarray}
	\lefteqn{\log\Big( \frac{M_{\rm BH}}{10^{8}\,h^{-1}\,M_\odot} \Big)\ = \ (-2.66\pm0.33) } \nonumber \\
	& & + \ (1.39\pm0.22)\,\log \Big [ \gamma^3\, H(z) \Big( \frac{M_{\rm vir}}{10^{13}\,h^{-1}\,M_\odot} \Big) \Big]~.
	\label{eqn:mass2bh}
\end{eqnarray}
This equation contains a redshift dependence through $H(z)$, something we will investigate later in Section~\ref{sec:BHhalo}.

Finally, we provide a few useful equations from the literature for converting to different optical filters.  First, to convert between bolometric quasar luminosity and $b_{\rm J}$-band absolute magnitude, \cite{Croom2005} provide
\begin{equation}
	M_{b_{\rm J}} = -2.66\,\log(L_Q)+79.42~,
\end{equation}
where $L_Q$ here is in watts and assumes $h=0.7$.  Second, to convert between $B$-band, $b_{\rm J}$-band (used by the 2dF QSO Redshift Survey, hereafter 2QZ), and $i$-band (the Sloan Digital Sky Survey Quasar Survey, hereafter SDSS) (k-corrected to $z\,=\,2$) quasar luminosities we use the conversions given in \cite{Croom2005} and \cite{Richards2006}
\begin{eqnarray}
	M_B = M_{b_{\rm J}} + 0.06 \nonumber ~,
\end{eqnarray}
\begin{equation}
	M_i[z=2] = M_{b_{\rm J}} - 0.71 ~.
	\label{eqn:convert}
\end{equation}

\subsection{Deciding which halos host quasars}
\label{sec:selection}

\begin{table*}
\centering
\caption{Parameters for the double power-law fit to the joint 2QZ and SDSS quasar luminosity functions at $z<3$ and $z \ge 3$ (Section~\protect\ref{sec:selection} and equations~\protect\ref{eqn:dblPL} and \protect\ref{eqn:Lstar}).  As in \protect\cite{Croom2004}, $h=0.7$ was assumed in the fit where relevant.}
\begin{tabular}{ccccccc}
\hline \hline
redshift range & $\alpha$ (bright slope) & $\beta$ (faint slope) & $M^*_{b_{\rm J}}(z=0)$ & $k_1$ & $k_2$ & $\Phi_* ({\rm Mpc}^{-3} {\rm mag}^{-1})$ \\
\hline \hline
$z<3.0$ & $-3.31$ & $-1.09$ & $-21.61$ & $1.39$ & $-0.29$ & $1.67 \times 10^{-6}$ \\
$z \ge 3.0$ & $-3.31+0.5\,(z-3)$ & $-1.09$ & $-21.61$ & $1.22$ & $-0.23$ & $1.67 \times 10^{-6}$ \\
\hline \hline
\end{tabular}
\label{tbl:parameters}
\end{table*}

Although Section~\ref{sec:mapping} provides a mapping between halo mass and quasar luminosity, not all halos will actually host a quasar at any given time.  To decide which do we use the quasar luminosity function.  

Specifically, we constrain the model to have the correct luminosity function at a given epoch using the technique of abundance matching (e.g. \citealt{Conroy2006}  and references therein).  First, we calculate the observed cumulative luminosity function, from bright to faint quasars, which provides a smoother representation of the data than the luminosity function alone.  Then, starting at the bright-end, we move faint-ward down the cumulative luminosity function in magnitude bins of $\Delta M=0.01$ and randomly dilute the number density of model quasars in the same magnitude bin to ensure they match the same cumulative abundance as the observations.  Hence, by construction, the model reproduces the observed quasar luminosity function.

The observed quasar luminosity function is only measured at discrete redshifts, and we would like to be able to build our model at any arbitrary redshift.  Hence, instead of using the data itself, we abundance match to a  functional fit of the data across multiple redshifts.  The two data sets we will later compare with are those of \cite{Croom2004} (2QZ), covering $0.4<z<2.1$, and \cite{Richards2006} (SDSS), covering $0.5<z<4.8$.

\cite{Croom2004} model their data using an evolving double power-law with the form
\begin{equation}
	\Phi(L_{b_{\rm J}},z) = \frac{\Phi(L^*_{b_{\rm J}})}{(L_{b_{\rm J}}/L^*_{b_{\rm J}})^{-\alpha} + (L_{b_{\rm J}}/L^*_{b_{\rm J}})^{-\beta}}~,
	\label{eqn:dblPL}
\end{equation}
where the characteristic luminosity, $L^*_{b_{\rm J}}$, is a function of redshift
\begin{equation}
	L^*_{b_{\rm J}}(z) = L^*_{b_{\rm J}}(z=0)\,10^{k_1z+k_2z^2}~.
	\label{eqn:Lstar}
\end{equation}
The authors used this functional form to fit the 2QZ quasar luminosity function out to $z \sim 2$.

Interestingly however, the \citeauthor{Croom2004} double power-law is a good fit to the quasar luminosity function well beyond its original fitting range.  This can be seen in figure~19 of \cite{Richards2006}, where the \citeauthor{Croom2004} result continues to provide a good match to the SDSS quasar luminosity function out to redshift three\footnote{Note that \citeauthor{Richards2006} fit their SDSS data using a simpler single power-law, which does not realistically represent the quasar faint-end.  Their data does not probe below $L^*$ at these redshifts (see however \citealt{Fontanot2007}).}.
Hence, to maintain the most realistic shape of the quasar luminosity function across the widest possible redshift range we adopt the fit of \cite{Croom2004} to $z=3$.  Above this redshift we modify their parameters slightly, adding evolution to bright-end power-law slope and softening the decline of $L^*_{b_{\rm J}}$ with redshift.  The fitting values we adopt are given in Table~\ref{tbl:parameters}.  The fits are shown below in Section~\ref{sec:QLF}.

\subsection{Model assumptions and parameters}
\label{sec:assumtions}

In order to keep our model as simple as possible while remaining accurate and (most importantly) understandable we list below our main simplifying assumptions:
\begin{itemize}
	\item Both the local $m_{\rm BH}-\sigma$ relation (equation~\ref{eqn:msigma}) and the local correlation between $\sigma$ and $v_c$ (equation~\ref{eqn:sigma}) exist at all redshifts and do not evolve with time.
	\item The relationship between halo virial velocity and galaxy circular velocity can be described by equation~\ref{eqn:vc} with $\gamma\,=\,1.0$ (see Appendix A5 of \citealt{Porciani2004} for a discussion on this point).
	\item Quasars can be modelled as a simple ``light bulb'', where, at any given time, they are either on or off.
	\item ``On'' quasars shine at half the Eddington luminosity (i.e. $\eta\,=\,0.5$ in equation~\ref{eqn:LQ}).  This is the mean observed value measured by \cite{McLure2004} during the height of quasar activity at $z=2$ (see also \citealt{Marconi2004}).
\end{itemize}
The impact of the uncertainty of these assumptions on our results is further discussed in Section~\ref{sec:errors}.

Although the above choices are reasonable they are not necessarily correct in detail.  For example, quasar luminosities are not either ``on'' or ``off'', but follow a light curve with a peak luminosity that is likely dependent on the specifics of the quasar trigger, the properties of the host galaxy, and redshift \citep[e.g.][]{Hopkins2006a}.  More complex modelling is required to capture these processes.  Our goal here is not to model the specifics of the quasar population in detail, but rather \textit{we aim to find the simplest model possible that can match a set of key observations, in spite of the missing detail.}  We will discuss this further in Section~\ref{sec:discussion}.

\section{Constraints, consequences and predictions to high redshift}
\label{sec:results}

\begin{figure}
\plotone{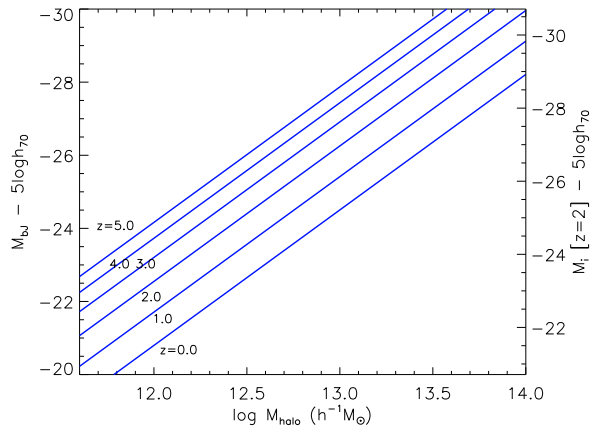}
\caption{The relationship between quasar luminosity and dark halo virial mass at various redshifts from $z=0$ to $z=5$ (equation~\protect\ref{eqn:mass2mag}).  Magnitudes are provided in both the $b_{\rm J}$-band used by the 2QZ survey and $i$-band used by the SDSS (k-corrected to $z=2$ -- see equation~\protect\ref{eqn:convert}).  Note that quasars of a fixed halo mass become brighter with increasing look-back time.}
\label{fig:mapping}
\end{figure}

\subsection{The quasar luminosity-halo mass relation}
\label{sec:LQmvir}

The relationship between quasar luminosity and dark matter virial mass is given by equation~\ref{eqn:mass2mag}.  This is plotted in figure~\ref{fig:mapping} for select redshifts out to $z=5$.  On the left axis we plot $b_{\rm J}$-band quasar absolute magnitude for comparison with the 2QZ survey results, and on the right axis we show $i$-band magnitudes for comparison with the SDSS quasar results (see equation~\ref{eqn:convert}).

Figure~\ref{fig:mapping} reveals that model quasars hosted by halos of a given mass get brighter the further back in time you look.  For example, in halos of $10^{12.5} M_\odot$, quasars brighten in luminosity by about two magnitudes between $z=0$ and $z=2$, and by another one-and-a-half magnitudes between $z=2$ and $z=5$.  One can see this by eye directly in the observational data, however, without deferring to theory or models.  The results of \cite{Croom2004} show similar amounts of brightening in the 2QZ quasar luminosity function between $z=0.4$ and $z=2.1$, while for the same catalogue, \cite{Croom2005} use clustering to infer that the masses of $L_*$ quasar dark halo hosts across the same redshift interval remain approximately constant at $10^{12-13}M_\odot$.  In this sense, our model is simply mimicking the data from which it was constrained.

From a theoretical point-of-view, the evolution in amplitude seen in figure~\ref{fig:mapping} originates from a redshift dependence in equation~\ref{eqn:vvir}, where, at fixed halo mass, the virial velocity of a halo increases with increasing redshift \citep[see also][]{Wyithe2003}.  This behaviour simply tells us that halos of equivalent mass live in higher $\sigma$-peaks at higher redshifts.

\subsection{The $\mathbf{m_{\rm{BH}}-\sigma}$ relation}
\label{sec:msigma}

\begin{figure}
\plotone{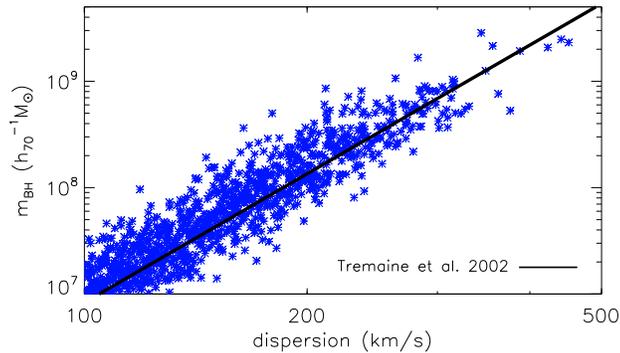}
\caption{The $m_{\rm BH}-\sigma$ relation between quasar black hole mass and galaxy velocity dispersion.  We assume in the model that this relationship is redshift independent (see Section~\protect\ref{sec:mapping}).  A fit to the observed local relation measured by \protect\cite{Tremaine2002} is shown by the solid thick line.}
\label{fig:msigma}
\end{figure}

The $m_{\rm{BH}}-\sigma$ relation is the primary observation used in our model to map quasar luminosity onto mass.  To produce a realistic quasar mock catalogue we include the observed $0.3\,{\rm dex}$ of scatter reported in \cite{Tremaine2002} when performing this mapping.  Our model relation is plotted in figure~\ref{fig:msigma}, where the points are a randomly selected sample of quasars from the model, and the thick solid line shows the best fit to the observed correlation.  It is important to note that, being a model quasar population, we are $100\%$ complete down to low dispersion, unlike the measured data.  Such measures, especially at low dispersion, still remain observationally formidable.

When constructing the model it was interesting to find that including the above dispersion does not appear important for its success in any way.  All results presented in this paper are essentially unchanged if we had rather simply made a direct mapping between $m_{\rm BH}$ and $\sigma$, ignoring the scatter.  Evidence is emerging to suggest that a tight correlation exists between $L_Q$ and $M_{\rm halo}$ \citep[e.g.][]{White2008}.  Understanding this result will be critical when using quasar clustering measures to observationally constrain the host masses of high redshift quasars.

We have made a critical assumption when constructing our model that the $m_{\rm{BH}}-\sigma$ does not evolve with redshift, either in amplitude or slope.  An evolving relation would place quasars of a given magnitude in either more massive halos (for a decreasing amplitude with increasing redshift) or less massive halos (for an increasing amplitude with increasing redshift).  This change will be reflected in the clustering properties of the quasars themselves.  As we show in Section~\ref{sec:clustering}, our non-evolving $m_{\rm{BH}}-\sigma$ relation assumption produces a quasar population whose 2-point function matches the observations extremely well out to at least $z\sim 4$.  The observational picture as to whether the $m_{\rm{BH}}-\sigma$ relation evolves remains unclear (see \citealt{Croton2006b} for a discussion), so for the current work we retain the non-evolving assumption.

\subsection{The quasar luminosity function to $\mathbf{z\sim 5}$}
\label{sec:QLF}

\begin{figure}
{\centering \leavevmode \psfig{file=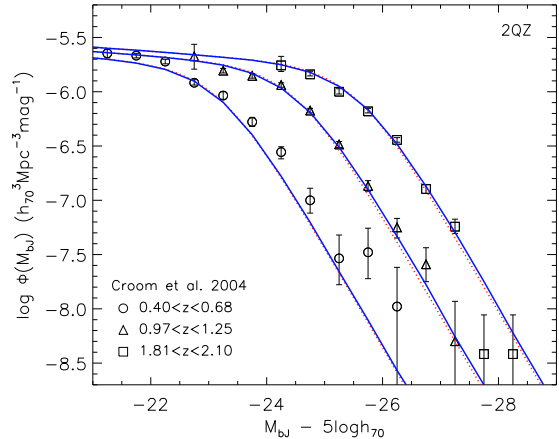,width=8.0cm,clip=}}
\caption{2QZ quasar luminosity functions \protect\citep{Croom2004} covering $z=0.40$ to $z=2.10$ (symbols with error-bars).  The (difficult to see) dotted lines show the fit to the data described by equation~\protect\ref{eqn:dblPL} with double power-law parameters given in table~\protect\ref{tbl:parameters}.  Solid lines show the model luminosity functions determined from abundance matching to the observations (Section~\protect\ref{sec:selection}).}
\label{fig:2QZLFs}
\end{figure}

As discussed in Section~\ref{sec:selection}, we use abundance mapping to select those halos with quasars that will produce a luminosity function identical to that observed at each redshift.  To maximise its versatility we match the model to a fit of the data that varies smoothly with redshift (Section~\ref{sec:selection} and table~\ref{tbl:parameters}), rather than to the data itself (which limits us to work at the observed redshifts only).

\begin{figure}
{\centering \leavevmode \psfig{file=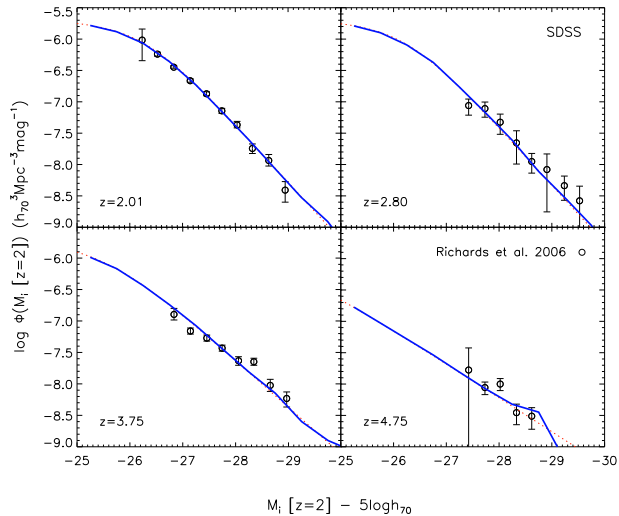,width=9.0cm,clip=}}
\caption{Similar to figure~\protect\ref{fig:2QZLFs}, but this time for quasars in the SDSS survey covering $z=2.01$ to $z=4.75$ \protect\citep{Richards2006}.  Again, dotted lines indicate the double power-law fit to the data (equation~\protect\ref{eqn:dblPL} and table~\protect\ref{tbl:parameters}), while the solid lines show the model quasar luminosity function (Section~\protect\ref{sec:selection}).}
\label{fig:SDSSLFs}
\end{figure}

Model luminosity functions are shown for $z<2$ in figure~\ref{fig:2QZLFs}, comparing with the 2QZ survey results of \cite{Croom2004}, and for $z>2$ in figure~\ref{fig:SDSSLFs}, comparing with the SDSS results of \cite{Richards2006}.  In each panel of both figures, the symbols show the observed data, the dotted line the double power-law fit to the data, and the solid line is the model.

It is no coincidence that the model line and double power-law fits are essentially indistinguishable for all but the lowest space densities at the highest redshifts (at which we are limited by the finite size of the Millennium Simulation).  Note that our extension to the \cite{Croom2004} double power-law representation of the 2QZ luminosity function (table~\ref{tbl:parameters}) provides a good fit to the data up to the highest redshifts probed by the SDSS quasar survey in figure~\ref{fig:SDSSLFs}.

\subsection{Quasar lifetimes}
\label{sec:lifetimes}

The quasar lifetime, $t_Q$, is defined here as $N_Q=t_Q/t_H$, where $N_Q$ is the fraction of halos at a given redshift that host a quasar, and $t_H$ is the Hubble time at that redshift.  $N_Q$ is a function of both mass and redshift and is determined in our model through abundance matching model quasar luminosities to the observed quasar luminosity function (Section~\ref{sec:selection}).  It is, in this context, a normalised quasar selection function.

\begin{figure}
\plotone{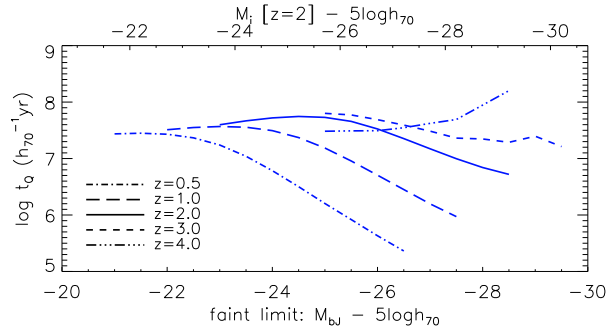}
\caption{Predictions for quasar lifetimes, $t_Q$, as a function of limiting faint quasar magnitude, at five different redshifts from $z=0.5$ to $z=4.0$.  
At $z\simlt 2$ and $L_Q>L^*$ the quasar lifetime shortens considerably.  At higher redshifts, $t_Q$ remains approximately constant or may even increase for the brightest.}
\label{fig:tQ}
\end{figure}

In figure~\ref{fig:tQ} we present quasar lifetimes as a function of limiting faint quasar magnitude for five redshifts, ranging from $z=0.5$ to $z=4$.  Lifetimes in general range from between $10^7$ to $10^8$ years.  At lower redshifts we see a distinct turn-over in $t_Q$, where brighter quasars have much shorter lifetimes (as short as $10^{5-6}$ years) than fainter quasars (levelling out at $\sim 10^{7.5}$ years).  This turn-over occurs at approximately $L^*$ in the quasar luminosity function (equation~\ref{eqn:Lstar}).  The turn-over is less pronounced for higher redshift quasars, flattening somewhat and even increasing to high luminosities at $z=4$.  

Due to the simplicity and transparency of the model we know exactly why model quasar lifetimes behave in this way: quasars occupy a narrower range of halo masses at late times relative to early times (skip forward to figure~\ref{fig:evolution} to see this).  Because of this, low redshift bright quasars are rare amongst the abundant $\sim 10^{13} M_\odot$ mass halos they occupy (hence the turnover at bright luminosities), whereas high redshift bright quasars are frequent among their rare $\sim 10^{13} M_\odot$ mass halo hosts (hence $t_Q$ remains flat).  Fainter quasars (those with $\sim L_Q<L^*$) tend to always commonly populate mostly abundant halos, again resulting in a relatively constant $t_Q$.  We will return to this point in Section~\ref{sec:application}.

Physically speaking, it is important to realise that the trends seen in figure~\ref{fig:tQ} do not result from the explicit modelling of a changing Eddington accretion fraction, as has often been explored \citep[e.g.][]{Hopkins2005d}.  Quasars in our model are assumed to always accrete at a fixed fraction of the Eddington rate across the quasar lifetime\footnote{Note that when $t_Q$ becomes longer than the doubling time for the black hole ($\simgt 10^8\, {\rm yr}$) black hole mass and quasar luminosity change non-trivially across the quasar lifetime.  This is an internal inconsistency that all light bulb models must navigate.  For our work, this is most relevant for extreme luminosity quasars at the highest redshifts probed.  Ignoring this effect for the sake of simplicity does not change our conclusions in any qualitative way.}.  
This is a key difference between our model and many previous works, and may explain its ability to simultaneously match such a wide range of observations.  For example, \cite{Wyithe2003} find an over-production of bright quasars at low redshift.  In our model, such quasars have very short lifetimes, and hence are not commonly seen in surveys.  This may simply arise due to the dwindling supply of cold gas in massive systems at late times \citep{Fabian1994}.

\subsection{The active and passive black hole mass functions}
\label{sec:BHMF}

Observations suggest that a black hole gains the majority of its mass while in the active high accretion (quasar) phase \citep{Heckman2004}.  The cumulative effect of such mass growth over cosmic time is measurable in the local (passive) black hole population.  Our model makes a prediction for the active black hole mass function.  It is important to note that this prediction arises via the dual constraint of linking quasar luminosity to halo virial mass through the $m_{\rm BH} - \sigma$ relation (equation~\ref{eqn:mass2mag}), and from matching the abundance of black holes to the quasar luminosity function (Section~\ref{sec:selection}) at various redshifts.  We did not tune the model in this regard to force a particular outcome.

\begin{figure}
{\centering \leavevmode \psfig{file=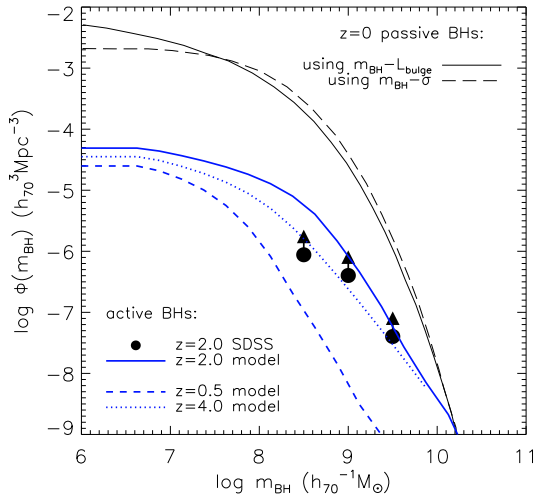,width=8.0cm,clip=}}
\caption{The cumulative black hole mass function.  The lower solid, dashed and dotted lines show the model result at three different redshifts, while the filled circles indicate lower limits measured from the SDSS $z=2$ quasar mass function of \protect\cite{McLure2004}.  Note that the model predicts a factor of $\sim 5$ more $10^{8.5}\,M_\odot$ black holes (quasars) than currently be seen in the data at $z=2$.  The upper solid and dashed lines show the local (passive) cumulative black hole mass function, calculated in two different ways.  The passive function is built from the continual production of quasars across cosmic time.}
\label{fig:BHMF}
\end{figure}

In figure~\ref{fig:BHMF} we show the observed cumulative black hole mass function for both $z=0$ passive and $z=2$ active black holes.  This figure is adopted from figure~6 of \cite{McLure2004}.  The upper thin solid and dashed lines show the local observed results inferred from the $m_{\rm BH}-$bulge luminosity relation and $m_{\rm BH}-\sigma$ relation, respectively.  
The three data points show the cumulative SDSS quasar mass function at $z=2$ for three different limiting black hole masses.  The upward pointing arrows on each indicate that the space density measured in each bin is incomplete, and hence provide only a lower limit to the true mass density.

The three thick lines in figure~\ref{fig:BHMF} show the model prediction for our complete sample of quasar black holes at redshifts $0.5$, $2.0$, and $4.0$, as indicated in the legend.  For black holes more massive than $\log m_{\rm BH}\simgt 9.0$ at $z=2$ our model is close to the lower limit found in the SDSS data.  If true, the model indicates (perhaps unsurprisingly) that much of the massive end of the black hole mass function forms solely from accretion during this time of peak activity. At lower black hole masses, $\log m_{\rm BH}\sim 8.5$, our quasar model predicts an excess of fainter quasars yet to be seen in the $z\sim2$ data.
At higher redshift, $z=4$, the model shows the continuing build-up of massive black holes but with less activity at lower masses.  By late times, $z=0.5$, massive black hole mass growth has largely stopped while the low mass holes continue to grow.  

This last behaviour is a black hole manifestation of the popular ``downsizing'' paradigm \citep{Heckman2004, Merloni2008}.  It is interesting that such shift in black hole mass growth with time arises naturally from the model constraints alone.  If we skip ahead to figure~\ref{fig:evolution} (Section~\ref{sec:application}) we can see the reason why.  Here, dashed lines show the changing space density of halos/quasars with time, while horizontal dotted lines show black hole mass.  At high redshift, very low space density contours are mostly flat (or only slowly rising) and correspond to a fixed black hole mass of $\sim 10^{9-10} M_\odot$.  At lower redshifts all density contours turn over sharply, and black hole mass decreases by up to a few orders-of-magnitude at fixed space density.  Hence, downsizing is predicted for all objects at $z \simlt 2$.   At $z \simgt 2$ the model predicts that low space density objects (i.e. the most massive) should show no downsizing trends (relative to $z=2$), while higher space density objects (i.e. closer to $L^*$) should be ``upsizing'', especially above redshifts $z \sim 4$.

\subsection{Evolution in the $\mathbf{m_{\rm{BH}}-M_{\rm halo}}$ relation and mass-to-light ratios}
\label{sec:BHhalo}

\begin{figure}
\plotone{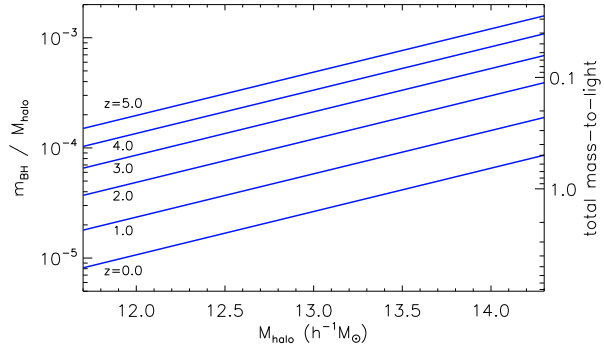}
\caption{Evolution in the $m_{\rm{BH}}-M_{\rm halo}$ relation and halo mass-to-bolometric luminosity ratio out to $z=5$ (equation~\protect\ref{eqn:mass2bh}).  There are two model predictions here: 1) that, at a give redshift, more massive halos should show an increasingly larger $m_{\rm{BH}}-M_{\rm halo}$ ratio (driven by accelerated black hole growth), and 2) the overall amplitude of this ratio should evolve with look-back time (see Section~\protect\ref{sec:BHhalo}).}
\label{fig:BHhalo}
\end{figure}

Section~\ref{sec:mapping} and equation~\ref{eqn:mass2bh} provide an analytic prediction for how black hole and dark matter halo mass are related.  This relationship includes a redshift dependence through $E(z) = [\Omega_{\rm m} (1+z)^3 + \Omega_\Lambda]^{1/2}$, implying that the $m_{\rm BH}-M_{\rm vir}$ ratio should evolve with time, with more massive black holes occupying dark matter halos of a fixed mass at higher redshifts relative to lower redshift.

In figure~\ref{fig:BHhalo} we use equation~\ref{eqn:mass2bh} to plot the hole-to-halo mass relation at five different epochs, from $z=0$ to $z=5$.  At any given redshift the ratio increases with increasing halo mass, implying that black holes become proportionally larger the more massive the halo is.  The evolution in the amplitude of $m_{\rm BH}$/$M_{\rm vir}$ with redshift can also be clearly seen, with the ratio changing by a factor of $\sim 5$ between $z=0$ and $z=2$, increasing to a factor of $20$ by $z=5$.  Previous authors have attempted to quantify the change in black hole to host galaxy/halo properties with time \citep[e.g.][]{Robertson2006, McLure2006, Croton2006b}.  Evolution of this type is a clear prediction of our model.  Our results are similar to those found by \cite{Wyithe2006} using different techniques \citep[see also][]{Wyithe2003}.

We can equivalently recast figure~\ref{fig:BHhalo} as a changing mass-to-light ratio using Equation~\ref{eqn:mass2mag}.  The right axis in figure~\ref{fig:BHhalo} shows this result.  Halos with masses greater than $10^{14}M_\odot$ host quasars with mass-to-light ratios less than unity regardless of the redshift of interest.  This is also true of all lower mass ($M_{\rm vir} \simlt 10^{13} M_\odot$) quasar/halo systems at redshifts $z\simgt 2$.

\begin{figure*}
{\centering \leavevmode \psfig{file=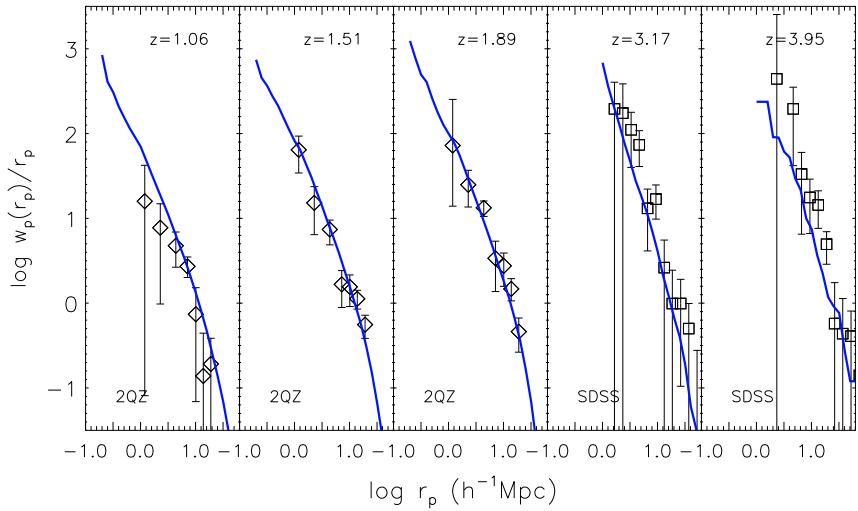,width=12cm}}
\caption{Model comparisons to the 2QZ \protect\citep{Porciani2004} and SDSS \protect\citep{Shen2007} projected redshift-space quasar correlation functions, from $z\sim1$ to $z\sim4$.  At all redshifts our model is a good match to the data (however with some small under-prediction of the clustering amplitude at $z>3$).  We also find a hint of steepening in the clustering on small scale scales ($r_p\simlt 1h^{-1}{\rm Mpc}$).}
\label{fig:CFs}
\end{figure*}

\subsection{Quasar clustering to $\mathbf{z\sim 4}$}
\label{sec:clustering}

The clustering of a given population of quasars will depend both on the masses of quasar hosts and the luminosity range which defines the quasar sample.  Both key observations used to constrain our quasar model, the $m_{\rm BH}-\sigma$ relation and quasar luminosity function, are relevant to shape the model 2-point function.

In figure~\ref{fig:CFs} we present the observed and model projected correlation functions at five discrete redshifts from $z=1$ to $z=4$.  The left three panels are taken from \cite{Porciani2004} using the 2QZ data, whereas the right two panels are those from \cite{Shen2007} with the SDSS data.  The marked redshift in each panel indicates the median redshift of the measured quasars.  For panels left-to-right, the absolute magnitude range defining each quasar sample is: the 2QZ survey $M_{b_{\rm J}}\!\in\![-25.32,-21.72]$, $M_{b_{\rm J}}\!\in\![-25.97,-22.80]$, and $M_{b_{\rm J}}\!\in\![-26.44, -23.37]$, and the SDSS survey\footnote{Note that \cite{Shen2007} do not state the absolute magnitude range that defines the quasars in their two high redshift bins.  Hence, for simplicity we select model quasars brighter than the faint absolute magnitude corresponding to $i=20.2$ (the SDSS quasar apparent magnitude limit) at the median redshift of each sample (M.~Strauss, priv. comm.).} $M_i\!<\!-26.1$ and $M_i\!<\!-26.7$ (all magnitudes have units of $5\log h_{70}$).

The default model produces a very good fit to the observed quasar clustering at all redshifts considered.  At $z>3$ there is a small under-prediction of the clustering amplitude, however the abundance of model quasars here is extremely low and the correlation function somewhat noisy.  This overall success gives us confidence that the use of the $m_{\rm BH}-\sigma$ relation and abundance mapping technique to construct our model is actually placing quasars in the correct halos, at least in a statistical sense.  At $z<2$ where the clustering is well measured, an upturn is seen at small scales ($r \simlt 1 {\rm Mpc}$).  Hints of such excess pairs have been found in the SDSS LRG-QSO cross correlation of \cite{Padmanabhan2008}, as well as work by \cite{Hennawi2006} and \cite{Myers2008}.  We leave a more detailed analysis of this result to future work.

\begin{figure}
\plotone{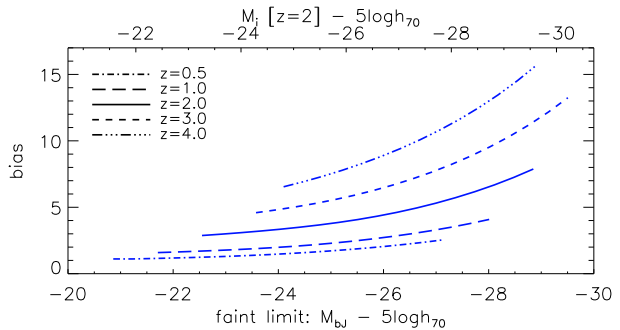}
\caption{Model predictions for the luminosity dependence of quasar bias, for quasar samples defined by a faint magnitude limit.  Results at five redshifts are shown, from $z=0.5$ to $z=4.0$.  Quasars at $z \simlt 1$ show little or no luminosity dependent bias.  At higher redshifts, a strong luminosity dependence in the bias is predicted for brighter quasars relative to faint. }
\label{fig:bias}
\end{figure}

\subsection{Luminosity dependent quasar bias at high redshift}
\label{sec:lumclustering}

For most redshift bins our statistics are far superior to that which can be measured observationally.  To test the model further we look for signatures of evolution in the clustering amplitude as a function of luminosity, captured here through the quasar bias.  Such evolution, or lack there of, is a more detailed probe of how quasars occupy halos, and is currently the focus of much observational scrutiny \citep{Porciani2006, Myers2007, daAngela2008}.

In figure~\ref{fig:bias} we show predictions for the luminosity dependent bias \citep{Mo1996} measured from our model, calculated using the \cite{Jenkins2001} mass function and a fit to the model quasar halo occupation fraction.  The calculation is performed analytically to remove the noise of small number statistics from rare objects at the high mass end.  Samples are defined with a faint limiting magnitude (i.e. applying equation~\ref{eqn:mass2mag} to a halo mass cut), and we present results for five redshift bins ranging from $z=0.5$ to $z=4$.

At $z \simlt 1$ weak (or no) luminosity dependent bias is present in the model quasar population.  At $z>1$, however, bright quasars show an increased bias with respect to faint quasars.  This ranges from $b \sim 3$ to $b \sim 7$ between luminosity extremes at $z=2$, and $b \sim 5$ to $b \sim 15$ at $z>2$.  Such increases are only marginally observable with current data, but will certainly be testable in future surveys as quasar numbers increase and the luminosity baseline widens.  

Both weak luminosity dependence at low redshift and significant luminosity dependence at high redshift constitute a firm prediction for the clustering of quasars in our model.  These predictions arise as a consequence of the changing occupation statistics of quasars in halos with time.  At high redshift quasars are spread over a much wider range of halo masses relative to low redshift for a comparable (to $L^*$) luminosity range (see figure~\ref{fig:evolution}).  This produces the stronger clustering gradient seen in figure~\ref{fig:bias}.  Said another way, \emph{luminosity dependent clustering evaporates at late times due to the narrowing of the range of halo masses that host quasars as the Universe ages}.  We will discuss this result further in Section~\ref{sec:application}.

\section{Discussion}
\label{sec:discussion}

\subsection{An (incomplete) overview of other popular quasar models}

Quasars have been modelled in a number of ways in the past several years \citep{Ciotti1997, Silk1998, Fabian1999, Kauffmann2000, Haehnelt2000, Cattaneo2001, Ciotti2001, Wyithe2002, Wyithe2003, Granato2004, Kawata2005, Begelman2005, Springel2005b, Springel2005c, DiMatteo2005, Hopkins2005a, Cattaneo2005, Cattaneo2005b, Hopkins2006a, Thacker2006, Lidz2006, Fontanot2006, Wyithe2006b, Malbon2007, Sijacki2007, Hopkins2008b, Hopkins2008a, Merloni2008, DiMatteo2008}.  The goal of such works has usually been to understand the cosmological evolution of quasars and their triggering mechanism and black hole gas accretion rates.  This is typically achieved by matching the model output to observables like the quasar luminosity function, the evolving space density of bright quasars, and the $m_{\rm BH}-\sigma$ relation.  More recently, quasars (and more generally AGN) have been linked to the quenching of star formation in massive elliptical galaxies, and quasar models have adapted to reflect these new found appreciations.

In fact, most modern AGN models are built upon the current belief that quasars are triggered by major merging events of gas rich galaxies \citep{DiMatteo2005, Hopkins2006a}.  This is a reasonable assumption to make.  Something significant must be happening to the gas in the galaxy to cause it to lose so much angular momentum, a necessary condition to drive gas into the central regions where the black hole resides.  Without such angular momentum loss it is hard to imagine how the required near Eddington accretion rates can be achieved.

From analytic arguments alone, \cite{Silk1998} postulate a critical black hole mass, determined by the surrounding halo properties, above which star formation has been suppressed due to an expanding quasar wind that sweeps the galaxy clean of its star forming gas.  Black holes in this picture either form early in the collapsing proto-galaxy at above the critical mass, or form close the the critical mass and are maintained at this mass by the hierarchical growth of the system.  This model provides an elegant explanation for many observed black hole -- galaxy/halo correlations but does not predict other quasar properties such as their luminosities or evolving space density.  Regardless, the \cite{Silk1998} model has become somewhat of a seed from which a number of more detailed models have developed.

One popular extension of these ideas is discussed in a series of papers by \citeauthor{Wyithe2002} \citep{Wyithe2002, Wyithe2003}.  In their model, quasars and their subsequent rapid black hole growth are triggered from major mergers, as discussed above.  Under the assumption that the local gas traps much of the quasar energy without radiating it away, and that the subsequent quasar luminosity is some fixed fraction of the Eddington luminosity, they derive a series of equations that allow them to predict quasar luminosity and black hole/host correlations at various redshifts.  With a small number of free (but physically motivated) parameters their model is tuned to provide a good fit to the high redshift quasar luminosity function, although it over-predicts the abundance of bright low redshift quasars. The model of \citeauthor{Wyithe2002} is somewhat similar to ours but differs in one critical way.  The abundance of quasars in their model is determined from halos who undergo rapid growth (i.e. major mergers).  Our modelling makes no such assumption, but rather forces the correct quasar number by abundance matching to the actual quasar luminosity function.  We discuss the advantages of this below.

\begin{figure*}
{\centering \leavevmode \psfig{file=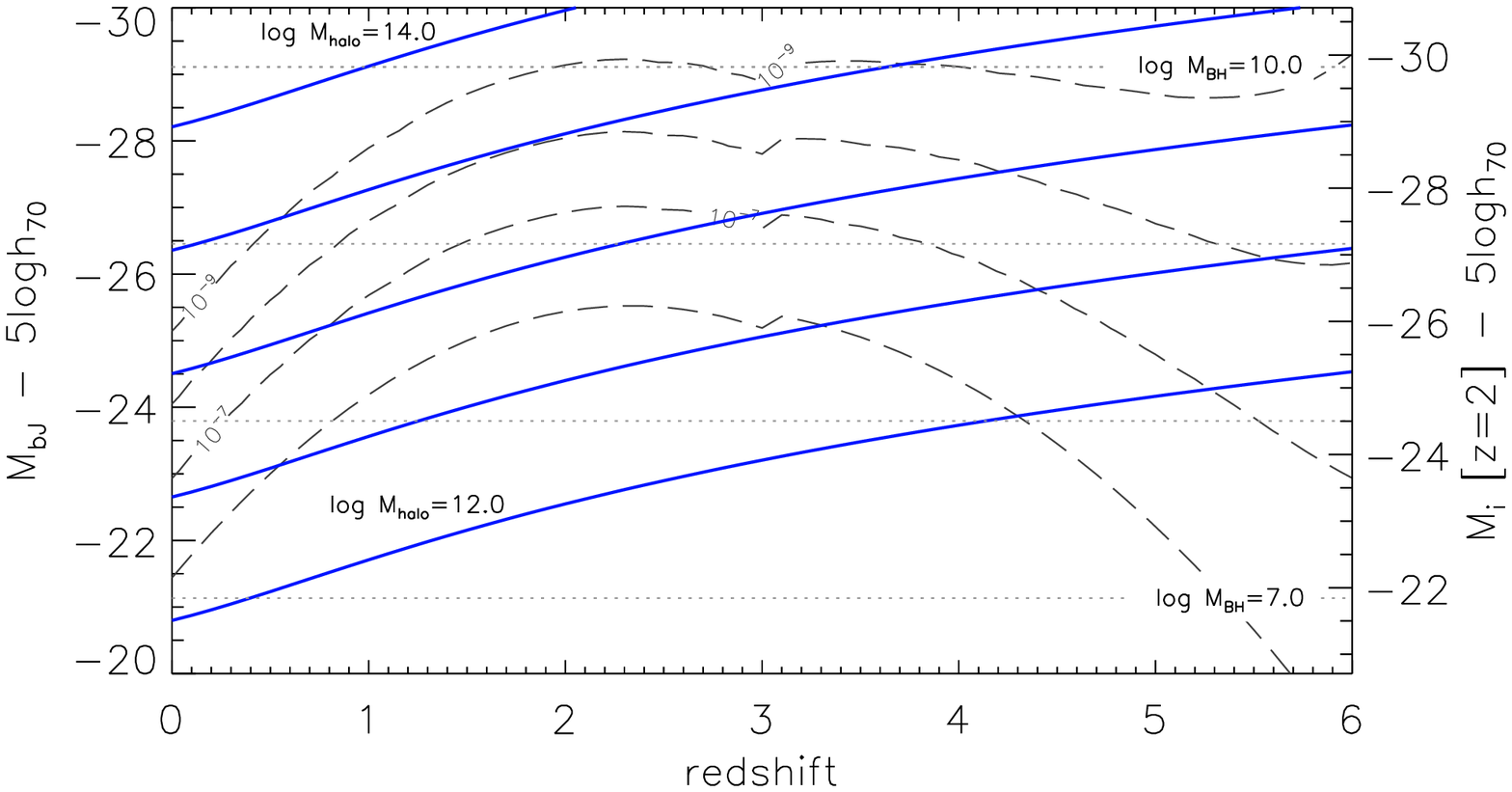,width=15cm}}
\caption{Halo virial mass predictions (solid lines) for quasars with measured luminosity and redshift, as defined by equation~\protect\ref{eqn:mag2mass}.  Magnitudes are provided in both the $b_{\rm J}$-band used by the 2QZ survey and $i$-band used by the SDSS (k-corrected to $z=2$ -- see equation~\protect\ref{eqn:convert}).  Dashed lines indicate the evolution of the host halo space density, as inferred from the observed quasar luminosity function.  These contours show that the distribution of mass for quasar host halos narrows as the Universe ages.  Horizontal dotted lines show the corresponding black hole mass, given by equation~\protect\ref{eqn:LQ}.  The downturn of density contours relative to black hole mass at low redshifts indicate downsizing in the black hole population.  Both upsizing and no-sizing are seen at higher redshifts, depending on the space density considered.}
\label{fig:evolution}
\end{figure*}

To further explore this picture of black hole and galaxy growth a number of authors have turned to performing hydrodynamic simulations of the complex merger and accretion processes themselves.  It should be remembered, however, that the physics of such processes cannot actually be resolved with current computing power, as the scales involved lie orders-of-magnitude below what is required.  Never-the-less, a combination of the phenomenological and hydrodynamic methodologies allow for an increased level of detail and accuracy in the models that is unreachable by analytic methods alone.

One example of this is the work by \citeauthor{Hopkins2006a}.  In a series of papers \citep{Hopkins2005a, Hopkins2005d, Hopkins2006a, Hopkins2006b, Hopkins2006c, Hopkins2007a, Hopkins2007b, Hopkins2008b, Hopkins2008a}, these authors explore the co-evolution of black holes and galaxies triggered by mergers using high resolution hydrodynamic simulations which include many realistic physical processes.  The most important aspect of their work is the inclusion of merger driven star formation and the redistribution of disk gas which drives both galactic bulge growth and, simultaneously, growth in the black hole through accretion.  Once convolved with cosmological statistics\footnote{\citeauthor{Hopkins2006a} simulate individual merger events, albeit a large number of them, and not the evolution of structure in a cosmological context.}, e.g. the evolution of merger rates with time, their model produces results that match a large number of observables: luminosity and mass functions, quasar lifetimes, Eddington ratios, and host galaxy properties.  

In a critical deviation from past models (including our work here), the simulations of \citeauthor{Hopkins2006a} explicitly track the varying black hole accretion rate during the merger.  This produces a range of predictions for how the quasar/AGN light curve should appear as a function of time, from first interaction to final merger remnant.  Their simulations suggest a possible common origin for the many observed AGN types, a so called ``unified'' merger driven model of AGN and galaxies \citep{Hopkins2006a}, that depends primarily on the time during the merger that the system is observed.  The beauty of their model is that it presents a well defined and testable picture of black hole and galaxy growth.  The drawback to their work is that its detail and complexity sometimes cloud its interpretation (see Section~\ref{sec:simple} below).

\subsection{A tool for observers}
\label{sec:application}

The quasar model presented in this paper provides a relationship between quasar luminosity, redshift, dark matter halo mass, and black hole mass.  The essence of this model is captured by equation~\ref{eqn:mass2mag} (Section~\ref{sec:eqns}).  For convenience, we invert this equation to obtain
\begin{eqnarray}
	\lefteqn{\log \Big( \frac{M_{\rm vir}}{10^{13}\,h^{-1}\,M_\odot} \Big) \ = \ (1.43\pm0.32) - \ \log\Big(\gamma^3\,H(z) \Big) } \nonumber \\
	& &  + \ (0.72\pm0.11)\, \log\Big( \frac{L_Q/\eta}{10^{12}\,h_{70}^{-1}\,L_\odot} \Big)~,
	\label{eqn:mag2mass}
\end{eqnarray}
where $H(z)$ is defined by equation~\ref{eqn:Hz}.  Equation~\ref{eqn:mag2mass} can be used to take a quasar with observed bolometric luminosity $L_Q$ at redshift $z$ and predict its halo virial mass, assuming values for parameters $\gamma$ (the relationship between virial and circular velocities) and $\eta$ (the Eddington luminosity fraction of the quasar).

Equation~\ref{eqn:mag2mass} is plotted in figure~\ref{fig:evolution}.  From this figure alone the quasar host virial mass (either individual or averaged over a group) for a given quasar luminosity and redshift may be simply read off (solid lines).  We emphasise, \emph{figure~\ref{fig:evolution} allows observers to determine statistically measured halo masses to complement their observations without the need for large quasar surveys or clustering measures.}  We discuss the uncertainty on such masses below.

Also over-plotted in figure~\ref{fig:evolution} are the corresponding mean space densities of quasar hosts (dashed lines), calculated from the quasar luminosity function\footnote{Note that the artificial bump at $z=3$ is due to a change in the assumed fitting parameters of the quasar luminosity function (table~\ref{tbl:parameters}) and has no bearing on the results.} (Section~\ref{sec:selection}), and the black hole mass at fixed luminosity  (horizontal dotted lines), as given by equation~\ref{eqn:LQ}.  

The spacing of density contours relative to halo mass highlights a main result of this work, which is that the distribution of the masses of dark matter halos hosting quasars narrows with decreasing redshift.  This leads to behaviour such as luminosity dependent clustering at high redshift but not at low (see Section~\ref{sec:lumclustering}), and luminosity dependent quasar lifetimes at low redshift but not high (see Section~\ref{sec:lifetimes}).  Also, as discussed Section~\ref{sec:BHMF}, the turnover of density contours at $z \simlt 2$ relative to black hole mass demonstrates downsizing in the black hole population.  At higher redshifts, rare massive black holes show no downsizing trend, whereas the more common $L^*$ quasars are predicted to be upsizing, especially at redshifts greater than $z \sim 4$.

\subsection{How well can halo mass be predicted?}
\label{sec:errors}

Before applying our model (particularly equation~\ref{eqn:mag2mass}) to an observation or set of observations it is prudent to understand the limits to which dark matter halo mass can be inferred given both the built in theoretical and observational uncertainties. 

Observational error is drawn from local measurements and has been propagated through each equation appropriately.  Near the characteristic luminosity of the quasar population, $L_Q/\eta\sim 10^{12}h_{70}^{-1}L_\odot$, the mass error has magnitude $0.32$ dex in log units.  One magnitude brighter or fainter than this increases the error to $0.34$ dex, whereas two magnitudes translates to an error of $0.39$ dex. While somewhat large at the extremes, this uncertainty is still sufficiently manageable that tight clustering constraints can be made and quasar host halo masses inferred from the model. 
 
Theoretically, our parameter $\gamma$ (describing the relationship between virial and circular velocities) can potentially take values of between $1.0$ and $1.8$, depending on halo mass and redshift. \cite{Porciani2004} argue that a value of unity is the most appropriate for high redshift quasars in the 2QZ survey, and hence is our choice. However, higher values may shift the inferred mass down by up to $0.5$ dex or more. According to \cite{Seljak2002}, such a shift would primarily affect quasars hosted by lower mass halos than considered here, $\sim 10^{11} h^{-1} M_\odot$.

Similarly, quasars are known to exhibit a range of Eddington values, somewhat in conflict with our single $\eta$ assumption (made to keep the model simple). However, \cite{McLure2004} argue that $\eta=0.5$ is a reasonable mean value for the high redshift quasar population. If we instead assume an $\eta$ value of $1.0$ we find a decrease in predicted halo mass by $\sim 0.2$ dex, while $\eta=0.1$ results in an increase of predicted halo mass of $\sim 0.5$ dex. 

Across the entire quasar population we believe our parameter choices are reasonable and justified (in the mean) by observation. However, individual quasars many not always be seen near their peak luminosity or have the virial-to-circular velocity ratios assumed here. Under these circumstances the mass inferred by our model will be incorrect. We emphasise that the quasar host halo masses predicted by our model are only accurate for the assumed $\eta$ and $\gamma$ values, and within the measured observational error. Relative to these assumptions they should provide a valuable tool with which to probe the quasar population across a wide luminosity and redshift range.

\subsection{Benefits to keeping it simple}
\label{sec:simple}

So what advantages does our model provide over past works?  First, and despite much circumstantial evidence (a lot of which is very convincing), astronomers still do not know the actual conditions and caveats under which quasar triggering occurs.  We may be wrong about the merger hypothesis, or the circumstances under which the triggering is otherwise (in)effective, or there may exist more than one mechanism to trigger quasars.  By using the data to determine which halos host quasars we free ourselves from pre-deciding the quasar trigger and the (perhaps unappreciated) consequences this may bring.  

Second, it is useful to build a statistically accurate representation of the quasar population (or at least one representation), from which we can `work backwards', so to speak.  Once our model is constructed we can examine it with confidence knowing that it has been constrained to be correct. A statistically correct model of the quasar population also allows observers to explore non-physics related issues in their data and survey design, such as cosmic variance and systematics.  

Third, our model is largely transparent in its cause and effect, which makes it easy for both theorists and observers to understand and apply.  This is rather a critical point.  One of the primary applications of any theoretical model is as tool to interpret the data in a physically meaningful way.  Unless it is clear \textit{why} a model behaves the way it does there is little insight to be gained from matching the data alone.  Of course, it is through the marriage of techniques that are both simple (to build intuition and set direction) and detailed (to understand the actual physics) that progress is made.

\section{Summary}
\label{sec:summary}

Under minimal assumptions we have demonstrated a (statistically) successful phenomenological method to occupy dark matter halos with quasars in a way consistent with many key observations out to $z\sim5$. We summarise the primary results and specific predictions of the model:
\begin{itemize}
\item We provide simple equations to predict host dark matter halo mass (equation~\ref{eqn:mag2mass}) and black hole mass (equation~\ref{eqn:mass2bh}) for quasars of given luminosity and redshift.  These equations are applicable to both single and group quasar observations.
\item We provide a new joint fit to the quasar luminosity functions of \cite{Croom2004} (2QZ) and \cite{Richards2006} (SDSS) which is accurate across $0.4<z<4.8$ (Section~\ref{sec:QLF} and figures~\ref{fig:2QZLFs}-\ref{fig:SDSSLFs}).
\item High redshift quasars of a given luminosity live in less massive dark matter halos than quasars of the same luminosity at low redshift.  Another way to say this is that, at fixed halo mass, high redshift quasars are brighter than their low redshift cousins (Section~\ref{sec:LQmvir} and figure~\ref{fig:mapping}).
\item Our model predicts luminosity dependent quasar lifetimes of $t_Q\sim 10^{7-8} h_{70}^{-1}\,{\rm yr}$, but which may be as short as $10^{5-6}\, {\rm yr}$ for quasars brighter than $L^*$ and $z \simlt2 $.  At $z \simgt 2$ this bright trend is less pronounced and even reverses (Section~\ref{sec:lifetimes} and figure~\ref{fig:tQ}).
This occurs because low redshift bright quasars are rare amongst the fairly abundant $\sim 10^{13} M_\odot$ mass halos that host them, whereas high redshift bright quasars are frequent among the rare $10^{13} M_\odot$ mass halos they occupy (figure~\ref{fig:evolution}).
\item Our active black hole mass function is consistent with that observed but predicts a significant excess (factor of $\sim 5$) of fainter quasars ($\log m_{\rm BH}\sim 8.5$) at $z\sim2$ (Section~\ref{sec:BHMF} and figure~\ref{fig:BHMF}). 
\item ``Downsizing'' naturally arises in our model; at fixed space density black hole mass decreases significantly at redshifts less than 2, while at higher redshifts this downsizing trend disappears for low space density objects and even reverses (i.e. ``upsizing'') for higher space density objects (figures~\ref{fig:BHMF} and \ref{fig:evolution}).
\item We predict evolution in the amplitude of the $m_{\rm BH}-M_{\rm halo}$ relation with time, with black holes of a given mass increasingly hosted by less massive halos at earlier times.  The amplitude changes by about a factor of $20$ between $z=5$ and $z=0$ (Section~\ref{sec:BHhalo} and figure~\ref{fig:BHhalo}).
\item Our model quasars have the correct clustering properties when compared to observations out to $z\sim 4$ (clustering was not a constraint on the model).  However, our clustering amplitude may be slightly low for quasars at $z\simgt 3$ (Section~\ref{sec:clustering} and figure~\ref{fig:CFs}).
\item Our model places quasars in halos in such a way that very little (or no) luminosity dependent clustering exists at $z\simlt 1$ for all magnitudes.  However, we predict strong luminosity dependent clustering at higher redshifts for luminous quasars when compared with the $L^*$ population (Section~\ref{sec:lumclustering} and figure~\ref{fig:bias}).  This behaviour results from the narrowing distribution of halo masses that quasars occupy as the Universe ages (figure~\ref{fig:evolution}).
\end{itemize}

Quasars appear set to remain a valuable probe of galaxy formation out to high redshift due to the distances they can cleanly be measured.  They are furthermore a incredibly interesting population of objects in their own right, encapsulating significant amounts of fundamental physics and broad phenomenology still yet to be understood.  The ability to produce statistically accurate models of this unique population will be essential to interpreting their future observation.

\section*{Acknowledgements}

The author would like to thank Sandy Faber, Joel Primack, and David Koo for encouraging me to return to this project after a long hiatus during a visit to UC Santa Cruz. Special thanks goes to Joe Hennnawi and Peder Norberg for valuable discussions during the `results stage' of this work.  Thanks as well to Carlton Baugh, Phil Hopkins, Peder Norberg, Yue Shen, and Martin White, and to the anonymous referee whose report improved the quality of this paper.

The author acknowledges support from NSF grant AST00-71048. The Millennium  Run simulation used in this paper was carried out by the Virgo Supercomputing Consortium at the Computing Centre of the Max Planck Society in Garching.  The halo catalogues used here are publicly available at http://www.g-vo.org/Millennium.  All mock quasar catalogues can similarly be found at http://astronomy.swin.edu.au/$\sim$dcroton.

\bibliographystyle{mnras}
\bibliography{../paper}

\label{lastpage}

\end{document}